\documentclass[runningheads]{llncs}
\usepackage[T1]{fontenc}
\usepackage{graphicx}
\usepackage{amsmath}
\usepackage{cleveref}
\usepackage{array}
\usepackage{multirow}

\newtheorem{pke}{Encryption Scheme}

\DeclareMathOperator{\ord}{ord}
\DeclareMathOperator{\dlog}{dlog}
\DeclareMathOperator{\RMPF}{RMPF}

\begin{document}
\title{Implementation of Learning with Errors in Non-Commuting Multiplicative Groups}
%
%

\author{Aleksejus Mihalkovich\orcidID{0000-0002-8661-3021} \and
Lina Dindiene\orcidID{0009-0004-5790-7495} \and
Eligijus Sakalauskas\orcidID{0000-0002-4620-4469}}
%
%
\institute{Kaunas University of Technology, Kaunas LT-44249, Lithuania \\
\email{\{aleksejus.michalkovic, lina.dindiene, eligijus.sakalauskas\}@ktu.lt}}
\maketitle              
\begin{abstract}

In this paper, we demonstrate a way to generalize learning with errors (LWE) to the family of so-called modular-maximal cyclic groups which are non-commuting. Since the group $\mathbf{M}_{2^t}$ has two cycles of maximal multiplicative order, we use this fact to construct an accurate criterion for restoring the message bit with overwhelming probability. Furthermore, we implement the original idea by O. Regev in the considered group to gain benefits from the non-commutativity of $\mathbf{M}_{2^t}$. Also we prove that using this approach we can achieve a level of security comparable to the original idea. 

\end{abstract}
\section{Introduction and Previous Work}

At the end of the XX century P. Shor in his paper \cite{Shor1994} proposed an algorithm for solving the discrete logarithm and integer factorization problems in polynomial time using quantum computations. Therefore, cryptographic primitives based on such problems are on the edge of extinction due to ongoing progress in the development of quantum processors. To our knowledge, as of September 2024, the record holders for most qubits are IBM with their 1121-qubit Condor \cite{IBM_Condor} and Atom Computing with their 1180-qubit computer \cite{Atom}. Moreover, IBM can offer 10 minutes per month of free access to their 127-qubit system to get a handle on quantum computing \cite{IBM_pricing}.

The threat of quantum cryptanalysis to the widely used classic algorithms like the RSA cryptosystem is so great that at the end of 2016, NIST announced a call for proposals for quantum-safe cryptographic algorithms \cite{NIST_cfp}. As of 2022, one public-key encryption and three digital signature schemes have been selected \cite{NIST_selected}.

Three out of four selected schemes use hard problems defined in lattices \cite{CRYSTALS-Kyber,CRYSTALS-Dilithium,Falcon}. The one we consider in this paper is the learning with errors (LWE) problem first introduced by M. Ajtai in his paper \cite{Ajtai1996}. This problem was originally formulated for $m$-dimensional lattice defined in the finite field $\mathbf{Z}_q$, where $q$ is prime, as follows:

\begin{definition}
	Let $\vec{a}$ be an $m$-dimensional vector with $a_i \in \mathbf{Z}_q$ and let $\chi$ be a probability distribution on $\mathbf{Z}$. The $LWE(\mathbf{Z}_q, m, \chi)$ problem is to recover a secret vector $\vec{x}$ given a pair $(\vec{a}, \vec{a} \cdot \vec{x} + \varepsilon)$, where $\vec{a} \in \mathbf{Z}_q^m$ is sampled uniformly and independently, and the error $\varepsilon \in \mathbf{Z}_q$ is sampled according to $\chi$.
\end{definition}

The basic idea behind LWE is using error terms added to linear equations. This composes a problem that is computationally infeasible to solve without revealing the secret vectors. O. Regev used this idea in his paper \cite{R2005} to define the following public key encryption scheme:

\begin{pke}\label{Regev_PKE} 
	Let $m$ and $n$ denote the number of equations and unknowns respectively. For a prime number $q \geq 2$ between $n^2$ and $2n^2$ let $\chi$ denote the probability distribution on $\mathbf{Z}_q$. Also, let $m = (1 + \epsilon)(n + 1) \log q$ for some constant $\epsilon > 0$. Then two entities called Alice and Bob, where Alice is the receiver and Bob is the sender, can interact by performing the following actions: 
	\begin{enumerate}
		\item Alice chooses a secret vector $\vec{x} \in \mathbf{Z}_q^n$ uniformly at random, where $n$ is the security parameter of the scheme. Her private key is $\vec{x}$.
		\item For $i = 1,\ldots, m$, Alice chooses $m$ vectors $\vec{a}_1, \vec{a}_2, \ldots, \vec{a}_m  \in \mathbf{Z}_q^m$ independently from the uniform distribution. Also she chooses elements $\varepsilon_1, \varepsilon_2, \ldots, \varepsilon_m \in \mathbf{Z}_q$ independently according to $\chi$. Her public key is given by $(A, \vec{b})$, where $A$ is an $m \times n$ matrix with columns $\vec{a}_1, \vec{a}_2, \ldots, \vec{a}_m$, and $\vec{b} = Ax + \vec{\varepsilon}$.
		\item To encrypt a bit, Bob chooses $k$ equations by picking a random set of row indices among all $2^m$ subsets of the set $\left[ m \right] = \{1, 2, \ldots, m\}$. We denote this set by $\mathbf{S} = \{i_1, i_2, \ldots, i_k\}$. Then Bob adds the chosen equations together to obtain an expression:
		\begin{equation}
			\Big(\sum_{i \in \mathbf{S}} a_{i1} \Big){x_1}+ \Big(\sum_{i \in \mathbf{S}} a_{i2}\Big){x_2} +  \ldots +  \Big(\sum_{i \in \mathbf{S}} a_{im}\Big){x_m} + \Big(\sum_{i \in \mathbf{S}} \varepsilon_i\Big)  =  \sum_{i \in \mathbf{S}} b_i.
		\end{equation}  
		If the original bit is $0$, the encryption is a pair $\Big(\vec{a}, b\Big)$, where $\vec{a} = \sum_{i \in \mathbf{S}} \vec{a}_i$ and $b = \sum_{i \in \mathbf{S}} b_i$. If the original bit is $1$, the encryption is $\Big(\vec{a}, \lfloor \frac{p}{2} \rfloor + b\Big)$.
		\item To decrypt a bit from the pair $(\vec{a}, b)$ Alice calculates $d = b - \vec{a} \cdot \vec{x}$, where $\cdot$ denotes the inner product of two vectors $\vec{a}$ and $\vec{x}$. Alice outputs $0$ if $d$ is closer to 0 than to $\frac{q}{2}$. Otherwise, she outputs $1$. 
	\end{enumerate}
\end{pke}

Analyzing the \Cref{Regev_PKE} we can see that the extra term $\lfloor \frac{q}{2} \rfloor$ used during encryption to encrypt $1$ is exactly the middle of the group $\mathbf{Z}_q$. Also note that during decryption the subtraction of the term $\vec{a} \cdot \vec{x}$ can always be performed since $\mathbf{Z}_q$ is a commuting group under addition. Furthermore, the decision Alice makes during the decryption step to restore the original bit relies on the properties of $\chi$, which is usually chosen to be the Gaussian distribution.    

Because LWE is conjectured to be intractable for quantum computers, this type of cryptography has become an important element in the search for a secure way for parties to communicate in today's digitized society. As a result over the recent years encryption-decryption scheme \cite{ZLZS2024,BHWA2022}, digital signature \cite{BGV2016}, and fully homomorphic encryption \cite{BV2011,BV2011C} were created. Moreover, LWE can also be used to provide zero-knowledge proof \cite{B2021,X2013}.

In 2010 Lyubashevsky et al. \cite{LPR2010} extended the LWE problem to ring-based structures. This extension paved the way for a new modified LWE method. Ring LWE (RLWE) was established to be more efficient in creating cryptographical primitives and developing lattice-based cryptosystems. Since it reduces the computational complexity by leveraging the structure of polynomial rings, ring LWE is the most significant and widely analysed modification to LWE \cite{LPR2013,O2021}.

Various modifications of algebraically structured LWE were introduced later \cite{PP2024}. Module learning with errors problem (MLWE) is a generalization of RLWE and was proven to be at least as hard as ring LWE problem \cite{A2017}. Another generalization of RLWE among others is order-LWE where the distribution of secret elements has a specific algebraic structure \cite{B2019}. Thus, the potential of LWE in cryptography is obvious and it seems that the possibilities of using this type of cryptographic primitives have not yet been exhausted.

Another branch of cryptography which is currently considered perspective is based on the study of hard problems defined in non-commuting groups. The first steps in this direction were made on the verge of the millennium, and so cryptographic primitives like Ko-Lee and Anshel-Anshel-Goldfield key exchange protocols were presented \cite{KL2000,AAG1999}.

In this paper, we also use the so-called right-sided matrix power function (RMPF) first introduced in \cite{SLT2008} as an example of the one-sided MPF. Formally, MPF is a mapping defined for a base matrix $W$, whose entries come from a multiplicative group $\mathbf{G}$ (abelian or otherwise), and takes either one (for one-sided MPFs) or two (for the two-sided MPF) scalar matrices $X$ and $Y$, whose entries come from a ring of integers $\mathbf{Z}_{\ord(\mathbf{G})}$, as an input \cite{SLT2008,SL2012,MSL2022}. Here the notation $\ord(\mathbf{G})$ stands for the multiplicative order of the group $\mathbf{G}$. The explicit definition of the right-sided MPF is as follows \cite{SLT2008}:

\begin{definition}
	Let $W$ be a $m \times n$ matrix with entries $w_{ij} \in \mathbf{G}$, where $\mathbf{G}$ is a multiplicative group, and let $X$ be an $n \times p$ matrix with entries $x_{ij} \in \mathbf{Z}_{\ord(\mathbf{G})}$. The right-sided MPF is a mapping $\mathbf{G}^{m \times n} \times \mathbf{Z}_{\ord(\mathbf{G})}^{n \times p} \rightarrow \mathbf{G}^{m \times p}$ denoted as
	\[
		\RMPF_W(X) = W^X,
	\]
	where the entries of matrix $V = \RMPF_W(X)$ are calculated as follows:
	\begin{equation}\label{RMPF}
		v_{ij} = \prod_{k = 1}^{n} w_{ik}^{x_{kj}}.
	\end{equation}
\end{definition}

The left-sided MPF can be defined in a similar way. Furthermore, the two-sided MPF can be defined as a combination of both one-sided MPFs \cite{SLT2008}. However, since these mappings are not used in this paper, we omit their explicit definitions. 

Since in this paper, the right-sided MPF is also used as an analog of the inner product for vectors $\vec{w} \in \mathbf{G}^m$ and $\vec{x} \in \mathbf{Z}_{\ord(\mathbf{G})}^m$ we introduce an extra notation $\vec{w}^{\vec{x}}$ to denote the following operation:
\begin{equation}\label{MPF_vec}
	\vec{w}^{\vec{x}} = \prod_{i = 1}^m w_i^{x_i}.
\end{equation}

Formally, according to the definition of the right-sided MPF, the vector $\vec{w}$ in \eqref{MPF_vec} should be transposed. However, since the upper index denoting transposing of $\vec{w}$ is inconvenient, we omit it in \eqref{MPF_vec} and onward.

Over the years several papers (see, for example \cite{MSL2020,MLDS2022,MLS2022}) have been published, where the authors have shown how MPFs can be applied in cryptography. Their results have shown that problems based on MPFs are hard and can potentially be quantum resistant \cite{S2012,SM2018}.

In this paper, we demonstrate how to expand the LWE problem to non-commuting groups therefore combining two perspective branches of cryptography together. We also present a security analysis of the presented schemes and compare them to the original Regev's idea. As usual, we present conclusions at the end of this paper.   

\section{A direct multiplicative approach to the generalization of LWE}

In this section, we present the simplest way to generalize LWE to multiplicative algebraic structures. Looking back at the \Cref{Regev_PKE}, we can make the following changes to adapt it for a multiplicative group $\mathbf{G}_q$:

\begin{pke}\label{direct_mult_PKE}
	Security parameters $m$ and $n$ are defined as in \Cref{Regev_PKE}. For a prime number $q \geq 2$ between $n^2$ and $2n^2$ let $\mathbf{G}_q$ be the Sylow subgroup of $\mathbf{Z}^\ast_p$, and let $\chi$ denote the probability distribution on $\mathbf{Z}_q$. 
	\begin{itemize}
		\item The coefficients $a_{ij}$ of the equations are chosen from $\mathbf{G}_q$ uniformly at random. Note that every element in $\mathbf{G}_q$ has a multiplicative order of a prime number $q | (p-1)$.
		\item The secret key is now a vector of exponents $\vec{x}$, where $x_i \in \mathbf{Z}_q$.
		\item The error terms $\varepsilon_i$ in the equations are generated in the form $\varepsilon_i = g^{r_i}$, where $g$ is a generator of $\mathbf{G}_q$ and $r_i \in \mathbf{Z}_q$ are chosen independently according to distribution $\chi$. Therefore, we have the following system of equations defined in $\mathbf{G}_q$:
		\begin{equation}\label{Sylow_LWE}
			\left\{
			\begin{array}{ccccccc}
				a_{11}^{x_1} & a_{12}^{x_2} & \ldots & a_{1m}^{x_m} & \varepsilon_1 & = & b_1 \\
				a_{21}^{x_1} & a_{22}^{x_2} & \ldots & a_{2m}^{x_m} & \varepsilon_2 & = & b_2 \\
				\ldots & \ldots & \ldots & \ldots & \ldots & \ldots & \ldots \\
				a_{m1}^{x_1} & a_{m2}^{x_2} & \ldots & a_{mm}^{x_m} & \varepsilon_m & = & b_m \\
			\end{array}
			\right.
		\end{equation}
		\item To encrypt a bit, we choose $k$ equations, denote the set of chosen rows by $\mathbf{S} = \{i_1, i_2, \ldots, i_k\}$, and multiply these equations in $\mathbf{G}_q$ to obtain the following result:
		\begin{equation}
			\Big(\prod_{i \in \mathbf{S}} a_{i1}^{x_1} \Big) \cdot \Big(\prod_{i \in \mathbf{S}} a_{i2}^{x_2}\Big) \cdot  \ldots \cdot  \Big(\prod_{i \in \mathbf{S}} a_{im}^{x_m}\Big) \cdot \Big(\prod_{i \in \mathbf{S}} \varepsilon_i\Big)  =  \prod_{i \in \mathbf{S}} b_i.
		\end{equation}
		\item Let us denote $a_j = \prod_{i \in \mathbf{S}} a_{ij}^{x_j}$, $b = \prod_{i \in \mathbf{S}} b_i$. If a bit is $0$, the encryption is $(\vec{a}, b)$, and if the bit is $1$, the encryption is $(\vec{a}, g^{\lfloor \frac{q}{2} \rfloor} b)$.
		\item To decrypt a bit, calculate $b(\vec{a}^{\vec{x}})^{-1} = g^r$, where $\vec{a}^{\vec{x}} = \prod_{i = 1}^{m}a_i^{x_i}$. Then we can make a decision based on the values of $g^r$. If $r$ is closer to 0 than to $\lfloor \frac{q}{2} \rfloor$ then the decrypted bit is $0$. Otherwise, it is $1$.  
	\end{itemize}
\end{pke}

However, this approach is of little interest to us, since it is basically the original idea written in terms of multiplicative notations. Furthermore, taking the discrete logarithm of all equations base $g$  we have the original LWE scheme defined in $\mathbf{Z}_q$. On the other hand, for this reason the presented scheme is as secure as the original idea.

Interestingly enough, the presented scheme can also be rewritten using matrix notation. Let us define:
\[
A = 
\begin{pmatrix}
	a_{11} & a_{12} & \ldots & a_{1m} \\
	a_{21} & a_{22} & \ldots & a_{2m} \\
	\ldots & \ldots & \ldots & \ldots \\
	a_{m1} & a_{m2} & \ldots & a_{mm}
\end{pmatrix},
\vec{x} = 
\begin{pmatrix}
	x_1 \\ x_2 \\ \ldots \\ x_m
\end{pmatrix},
\vec{\varepsilon} = 
\begin{pmatrix}
	\varepsilon_1 \\ \varepsilon_2 \\ \ldots \\ \varepsilon_m
\end{pmatrix},
\vec{b} = 
\begin{pmatrix}
	b_1 \\ b_2 \\ \ldots \\ b_m
\end{pmatrix}.
\]

Then we can rewrite system \eqref{Sylow_LWE} as follows:
\begin{equation}\label{Sylow_LWE_matrix}
	A^{\vec{x}} \odot \vec{\varepsilon} = \vec{b},
\end{equation}
where $A^{\vec{x}}$ is the right-sided MPF defined by \eqref{RMPF} and $\odot$ is the Hadamard product of two vectors.

We can draw inspiration from a couple of observations in the \Cref{direct_mult_PKE} to adapt the original idea to another multiplicative group. First, we can see that we used only a certain subgroup of a larger multiplicative group $\mathbf{Z}^\ast_p$ to define the \Cref{direct_mult_PKE}. This was done to ensure that every element of the Sylow subgroup $\mathbf{G}_q$ aside from $1$ has a multiplicative order of $q$. Therefore, the private key exponents are chosen from $\mathbf{Z}_q$. Also note that the multiplicative algebraic structure to define the \Cref{direct_mult_PKE} (in our case, $\mathbf{G}_q$) must be a group, since otherwise the multiplicative inverse may not exist and hence the decryption step fails. Furthermore, the decision on the decrypted bit is more complicated as compared to the \Cref{Regev_PKE} since it is harder to consider the leftover term $g^r$. This issue may be resolved by making a list of values for which the decrypted bit is 0, and for which the decrypted bit is 1. Weather this is a reasonable solution depends on the value of $q$. Another simple solution is to calculate the discrete logarithm of $g^r$ base $g$, i.e. $r = \dlog_g g^r$. In this case, the decryption works just as in the Regev's original idea.

\section{Description of the multiplicative group}

In our work, we mainly consider multiplicative groups. Recently our attention turned to the study of non-commuting groups in cryptography, since problems defined in such structures are generally hard to solve and have the potential to withstand quantum analysis.

In this paper, we focus on one specific family of modular-maximal cyclic groups of order $2^t$ generally denoted as $\mathbf{M}_{2^t}$ \cite{GS_ord16,GS_ord32,GS_ord64}, where $t \geq 3$ is the size-defining parameter. The general representation of these groups is presented below:
\begin{equation}
	\mathbf{M}_{2^t} = \langle a, b | a^{2^{t-1}} = e, b^2 = e, ab = ba^{2^{t-2}+1}\rangle,
\end{equation}
where $a$ and $b$ are the generators and $e$ is the identity of the group.  We can see from the definition of $\mathbf{M}_{2^t}$ that the generators $a$ and $b$ do not commute if $t \geq 3$. Furthermore, due to the third relation, any element of the form $a^{k_1} b^\alpha$ can be represented in the form $b^\alpha a^{k_2}$ (here $\alpha \in \{0, 1\}$ and $k_{1,2} \in \{0, 1, \ldots, 2^{t-1} - 1\}$). Throughout this paper we use the form $b^\alpha a^k$ to represent the elements of $\mathbf{M}_{2^t}$.

Originally the group $\mathbf{M}_{16}$ caught our attention since it is one of seven non-commuting groups of size $16$ which cannot be decomposed into smaller ones \cite{GS_ord16}. Furthermore, similar results have been proven for groups $\mathbf{M}_{32}$ and $\mathbf{M}_{64}$ \cite{GS_ord32,GS_ord64}. We think that this feature beneficially distinguishes these groups from commuting ones, since most of the abelian groups can be split into either a direct or a free product of smaller groups. Therefore, for commuting multiplicative groups their generators can be considered separately, which results in more LWE equations for the attacker to study.

As we have seen from the presented direct multiplicative approach given by \Cref{direct_mult_PKE}, to construct an LWE problem using a multiplicative group we need the multiplication and exponentiation operations. Also we should be able to calculate the multiplicative inverse of an element. Let us now present explicit expressions of these operations.

First, we consider the multiplication of two elements $\omega_1 = b^{\alpha_1}a^{k_1}$ and $\omega_2 = b^{\alpha_2}a^{k_2}$ in $\mathbf{M}_{2^t}$. We have:
\begin{equation}\label{m16mult}
	\omega_1 \cdot \omega_2 = 
	\left\{
		\begin{array}{c} 
			b^{\alpha_1 +\alpha_2} a^{k_1 +k_2} \text{, if } k_1 \text{ is even;} \\ 
			b^{\alpha_1} a^{k_1 +k_2} \text{, if } k_1 \text{ is odd and } \alpha_2 = 0; \\
			b^{\alpha_1 +1} a^{k_1 +k_2 + 2^{t-2}} \text{, if } k_1 \text{ is odd and } \alpha_2 = 1.
		\end{array}
	\right.
\end{equation}

The expression for raising the element $\omega = b^\alpha a^k$ to a power $n$ can be derived from \eqref{m16mult} and is as follows:

\begin{equation}\label{m16exp}
	\omega^n =
	\left\{
		\begin{array}{c} 
			a^{kn} \text{, if } \alpha = 0; \\
			b^{n} a^{kn} \text{, if } \alpha = 1 \text{ and } k \text{ is even;} \\
			b^{n} a^{kn+2^{t-2}\left[\frac{n}{2} \right]} \text{, if } \alpha = 1 \text{ and } k \text{ is odd.} 
		\end{array}
	\right.
\end{equation}

Immediately we see from \eqref{m16exp} that the maximal multiplicative order of the elements in $\mathbf{M}_{2^t}$ is equal to $2^{t-1}$. Hence any power $n$ in \eqref{m16exp} can be considered modulo $2^{t-1}$. This is one of the key facts we use in this paper.

Finally, the inverse of the element $\omega = b^\alpha a^k$ can be calculated as:

\begin{equation} \label{m16inv} 
	\omega^{-1} =
	\left\{
		\begin{array}{c} 
			a^{-k} \text{, if } \alpha = 0; \\ 
			ba^{-k} \text{, if } \alpha = 1 \text{ and } k \text{ is even;} \\
			ba^{2^{t-2}-k} \text{, if } \alpha = 1 \text{ and } k \text{ is odd.} 
		\end{array}
	\right.  
\end{equation}

This expression can also be derived from \eqref{m16exp} by calculating $\omega^{2^{t-1} - 1}$.

All of the presented formulas can be easily proven using the definition of $\mathbf{M}_{2^t}$. For the case of $t = 4$ these proofs can be found in \cite{MSL2020}. 

In our paper, we also make use of the following important observation: an extra summand of $2^{t-2}$ appears only in the third case of expressions \eqref{m16mult} and \eqref{m16exp}. We use this fact to help us control the appearance of extra factors $a^{2^{t-2}}$ when defining the LWE scheme in $\mathbf{M}_{2^t}$. More precisely, setting $f(n) = 2^{t-2}\left[\frac{n}{2} \right] \bmod 2^{t-1}$ we can see from \Cref{function} when the extra summand appears in \eqref{m16exp}.
\begin{table}[h]
	\caption{Values of function $f(n)$ \cite{MSL2020}.}\label{function}
	\centering
	\begin{tabular}{|p{0.3in}|p{0.3in}|p{0.3in}|p{0.3in}|p{0.3in}|p{0.3in}|p{0.3in}|p{0.3in}|p{0.3in}|p{0.3in}|p{0.5in}|p{0.5in}|p{0.5in}|p{0.5in}|} \hline
		
		$n$ & 0 & 1 & 2 & 3 & 4 & 5 & 6 & 7 & \ldots & $2^{t-1} - 4$ & $2^{t-1} - 3$ & $2^{t-1} - 2$ & $2^{t-1} - 1$ \\ \hline 
		$f(n)$ & 0 & 0 & $2^{t-2}$ & $2^{t-2}$ & 0 & 0 & $2^{t-2}$ & $2^{t-2}$ & \ldots & 0 & 0 & $2^{t-2}$ & $2^{t-2}$ \\ \hline 
	\end{tabular}
\end{table}

Despite the word 'cyclic' in its name, the $\mathbf{M}_{2^t}$ does not contain any single element, which generates the whole group. However, the are two cycles of order $2^{t-1}$. These are presented below:
\begin{gather*}
	\langle a \rangle = \{e, a, a^2, \ldots, a^{2^{t-1} - 1}\}; \\
	\langle ba \rangle = \{e, ba, a^2, \ldots, ba^{2^{t-1} - 1}\}.
\end{gather*}

It is easy to see that the presented cycles intersect at even powers of generator $a$, which define the center of $\mathbf{M}_{2^t}$. Also, the presented cycles contain only $75\%$ of total $2^t$ elements of the considered group.

Notably there are other cycles of smaller orders in $\mathbf{M}_{2^t}$. These are less interesting to us, since we desire to make use of the non-commutativity of $\mathbf{M}_{2^t}$.

Before we end this section, let us present the following proposition which is used in this paper to construct a working scheme:

\begin{proposition}
	For any two elements $\omega_1, \omega_2 \in \mathbf{M}_{2^t}$ and $n \in \mathbf{Z}_{2^{t-1}}$ we have:
	\begin{equation}\label{m16distr}
		(\omega_1 \omega_2)^n =  a^{\alpha 2^{t-2}} \omega_1^n \omega_2^n,
	\end{equation}
	where $\alpha \in \{0, 1\}$.
\end{proposition}

\begin{proof}
	This fact follows from the definitions \eqref{m16mult} and \eqref{m16exp} of basic operations in group $\mathbf{M}_{2^t}$.
\end{proof}

Note that in \eqref{m16distr} we have $\alpha = 0$, if both $\omega_1$ and $\omega_2$ come from the same cycle. This is one of the facts, which helps us to control additional factors during the decryption step. 

\section{LWE in the non-commuting multiplicative group}

Let us now present the main idea of this paper, i.e. we show how LWE can be adapted to work in a non-commuting group $\mathbf{M}_{2^t}$. Inspired by the original idea of Regev we propose to divide the cycle $\langle ba \rangle$ into two equal parts. However, we also need several other changes to our proposal to achieve the benefits from a non-commuting group while working properly.

\begin{pke}\label{m16_2c_PKE}
	We assume that the number of equations $m$ and the number of unknowns $n$ are fixed. Then we can define the following one-bit encryption scheme based on LWE in $\mathbf{M}_{2^t}$ as follows:
	
	\begin{enumerate}
		\item Alice generates her private key as a vector $\vec{x}$, where $x_i \in \mathbf{Z}_{2^{t-1}}$.
		\item She randomly picks an integer $n_c \in \{1, 2, \ldots m - 1\}$. Then Alice uniformly and independently chooses entries of the base matrix $W$ from $\mathbf{M}_{16}$, so that the first $n_c$ columns of $W$ contain elements from the cycle $\langle ba \rangle$, and the leftover columns contain elements from $\langle a \rangle$. Therefore, entries in all the columns of matrix $W$ commute whereas row entries are divided into two commuting parts.
		\item Alice now calculates the vector $\vec{u} = W^{\vec{x}}$ and samples the errors $\varepsilon_i$ from the discrete Gaussian distribution $\mathcal{N}(0, \sigma)$, where $\sigma \leq 2^{\frac{t-1}{4}}$. Alice publishes the pair $(W, \vec{v})$ as her public key, where $\vec{v} = \vec{\varepsilon} \cdot \vec{u}$.
		\item To encrypt a bit, Bob chooses a random subset $\mathbf{S} = \{s(1), s(2), \ldots, s(r)\}$ of $\left[ m \right]$. Also, for each of the chosen equations he sample error terms $\delta_{s(i)} \in \{e, ba\}$ as follows:
		\begin{itemize}
			\item For the first of the chosen equations he chooses an error $\delta_{s(1)}$ so that the product $v^\prime_{s(1)} = \delta_{s(1)} v_{s(1)}$ is of the form $ba^k$ for some $k$, i.e. the generator $b$ is present;
			\item For all of the subsequent chosen equations Bob chooses errors $\delta_{s(i)}$ so that the product $v^\prime_{s(i)} = \delta_{s(i)} v_{s(i)} \in \langle a \rangle$, i.e. the generator $b$ is absent.
		\end{itemize}
		\item He calculates $\vec{w}$, where $w_j = \prod_{i \in \mathbf{S}}a_{ij}$, and $v^\prime = \prod_{i = 1}^{r} v^\prime_{s(i)}$. If the original bit is $0$ Bob outputs $(\vec{w}, v^\prime)$, otherwise the encryption is $(\vec{w}, a^{2^{t-2}} v^\prime)$.
		\item To decrypt a bit, whose encryption is a pair $(\vec{w}, v^\prime)$, Alice uses her private key $\vec{x}$ to calculate $h = v^\prime \cdot (\vec{w}^{\vec{x}})^{-1}$. Since $h \in \langle ba \rangle$, Alice can calculate the discrete logarithm of $h$ base $ba$. The decrypted bit is $0$ if $\dlog_{ba} h$ is closer to 0 than to $2^{t-2}$ and 1 otherwise.
	\end{enumerate}
\end{pke}

Evidently, the non-commutativity of $\mathbf{M}_{2^t}$ cannot be ignored in this case. Moreover, the two columns can only be switched if they commute. Otherwise, it becomes hard to control the extra factors. 

Another major difference is the fact that Bob does not use the vector $\vec{v}$ directly to produce the ciphertext of a bit. In other words, the \Cref{m16_2c_PKE} does not work correctly if the errors are absent unless the entries of $\vec{u}$ are contained in $\langle a \rangle$ in the first place. Simply put, the error term in the LWE problem produced by \Cref{m16_2c_PKE} always exists.

Now we prove the correctness of our proposal, i.e.

\begin{proposition}\label{prop:main_dec}
	The \Cref{m16_2c_PKE} decrypts the encrypted bits correctly with overwhelming probability.
\end{proposition}

  \begin{proof} 
We begin the proof by analyzing an equation computed by Alice in \Cref{m16_2c_PKE}. Let us denote the result of the  $k$-th equation by $v_k$:
\begin{equation}
	v_k=\varepsilon_k \cdot(ba)^{\alpha_k}(a)^{\gamma_k}=(ba)^{\beta_k}(ba)^{\alpha_k}(a)^{\gamma_k},
\end{equation}
where $\beta_k$ is distributed according to the discrete Gaussian distribution $\mathcal{D}_{\mathbf{Z},\sigma}$; $\alpha_k,{\gamma_k}$ are uniformly distributed over $\mathbf{Z}_\rho$, where we denote $\rho = 2^{t-1}$ for simplicity, and all three variables are independent. It is straightforward to verify that the distribution of $v_k$ is uniform. In fact, it is the basis of the proof of \Cref{prop:main_sec}. Moreover, for all $t_0,t_1 \in \mathbf{Z}_\rho$ we have:

\begin{eqnarray}\label{alphabeta}
	&& \hskip -1cm \Pr\big[(ba)^{\beta_k+\alpha_k}=(ba)^{t_0},(ba)^{\beta_k}=(ba)^{t_1}\big] \nonumber\\ &=& \Pr\big[(ba)^{\alpha_k}=(ba)^{t_0-t_1},(ba)^{\beta_k}=(ba)^{t_1}\big] \nonumber \\ &=& \frac{1}{p} \Pr\big[(ba)^{\beta_k}=(ba)^{t_1}\big],
\end{eqnarray}
i.e., the variables $(ba)^{\beta_k+\alpha_k}$ and $(ba)^{\beta_k}$ theoretically are independent; in the last step, we utilized the independence of $\alpha_k$ and $\beta_k$.

Let us now turn to Bob’s calculations. The result of the $k$-th equation is as follows:
\begin{equation}
	v^\prime_{k}=\delta_k \cdot \epsilon_k \cdot (ba)^{\alpha_k}(a)^{\gamma_k}=(ba)^{\tau_k}(ba)^{\beta_k+\alpha_k}(a)^{\gamma_k},
\end{equation}
where $\tau_k$ is distributed depending on the outcome of \(\alpha_k+\beta_k\):
\[
(ba)^{\tau_k} = 
\begin{cases} 
	(ba)^0 & \text{if } (ba)^{\alpha_k+\beta_k}\in \langle a \rangle \\
	(ba)^1 & \text{if } (ba)^{\alpha_k+\beta_k}\in \langle ba \rangle.
\end{cases}  
\]

Note that for the first equation these cases are switched. However, this fact is irrelevant in this proof.

Based on the formation of Alice’s equation $u$ and the properties of the multiplicative group $\mathbf{M}_{2^t}$, we observe that $ (ba)^{\alpha_k+\beta_k}$
includes the generator $b$ in the result in half of the cases, and excludes it in the other half. Thus, we see that $(ba)^{\tau_k}$
is a Bernoulli random variable with parameter $p=\frac{1}{2}$, $(ba)^{\tau_k} \sim \mathcal{B}(\frac{1}{2})$, that depends on the outcome of $(ba)^{\alpha_k+\beta_k}$. It should be noted that $(ba)^{\tau_k}$  depends on the final result of $(ba)^{\alpha_k+\beta_k}$, but does not depend directly on $\alpha_k$ or $\beta_k$, because the uniform distribution of 
$(ba)^{\alpha_k+\beta_k}$ theoretically does not depend on the value of 
$\beta_k$, provided that the distribution of 
$\alpha_k$ is uniform (as shown in \eqref{alphabeta}).

By multiplying $r$ equations randomly selected by Bob, we obtain $v^\prime$:
\begin{equation*}
	v^\prime =\prod_{i=1}^r \delta_i \prod_{i=1}^r \epsilon_i \prod_{i=1}^r (ba)^{\alpha_i} \prod_{i=1}^r (a)^{\gamma_i}.
\end{equation*}

Based on \Cref{m16_2c_PKE}, to evaluate the correctness of decryption, it remains to analyze only the modified error term $h = \prod_{i=1}^r \delta_i \prod_{i=1}^r \varepsilon_i$.

The distribution of $\prod_{i=1}^r \delta_i = (ba)^{\tau}$ is binomial with parameters $r$ and $\frac{1}{2}$, since each marginal distribution 
of $\delta_i$ is $\mathcal{B}(\frac{1}{2})$ and all $\delta_i$ are independent.

The product of $r$ independent discrete Gaussian random variables of the form $\varepsilon_i=(ba)^{\beta_i}$ implies that the exponent of the term $(ba)^{\beta}$, where $\beta = \sum_i^r\beta_i$ is itself a discrete Gaussian random variable, whose distribution can be readily written down. Since $\delta_i = (ba)^{\tau_i}$ does not directly depend on $\varepsilon=(ba)^{\beta_i}$, the multivariate distribution of the exponent of the modified error term $(ba)^{\tau+\beta}$ is as follows:
\begin{eqnarray}\label{prob1}
	&& \hskip -1cm
	\Pr\left(\tau+\beta = t\right) = \Pr\left(\tau+\sum_{i=1}^r \beta_i = t\right) \nonumber \\ &=& \sum_{\substack{t_1, t_2, \dots, t_n \in \mathbf{Z}_p \\ \tau \in \{0,1,\dots,r\} \\ t_0+ t_1 + t_2 + \cdots + t_n = t}} \Pr(\tau=t_0) \prod_{i=1}^n \Pr(\beta_i = t_i)
	={P}_{\tau+\beta}(t),
\end{eqnarray}
for each $t \in \mathbf{Z}_\rho$. This is subsequently used to evaluate the probability of error in the decryption step (the final step performed by Alice in \Cref{m16_2c_PKE}). This distribution reflects the probability of correctly decrypting $0$, if the resulting value of the exponent $\tau+\beta$ is closer to $0$ than to $2^{t-2}$, i.e. $\tau+\beta \in [0; \frac{2^{t-2}}{4}] $ or $\tau+\beta \ge  \frac{(3\cdot 2^{t-2})}{4} $. The shifted distribution corresponds to the probability of correctly decrypting 1, i.e. the modified error distribution is shifted by $\frac{2^{t-2}}{2}$:
\begin{eqnarray}\label{prob2}
	\Pr\left(\tau+\beta +\frac{2^{t-2}}{2} = t\right) ={P}_{\tau+\beta}\Big(t-\frac{2^{t-2}}{2}\Big),
\end{eqnarray}
and decryption is correct if the resulting value of $\tau+\beta$ remains within the interval $\Big(\frac{2^{t-2}}{4};\frac{(3\cdot 2^{t-2})}{4} \Big)$.

Hence, the decryption fails if $ \tau+\beta \in \Big(\frac{2^{t-2}}{4};\frac{(3\cdot 2^{t-2})}{4} \Big)$ when the encrypted bit is $0$, or if $\tau+\beta \in [0; \frac{2^{t-2}}{4}] $ or $\tau+\beta \ge  \frac{(3\cdot 2^{t-2})}{4} $ when the encrypted bit is $1$.

At this point, the decryption failure probability can be written in terms of the distributions given in equations \eqref{prob1} and \eqref{prob2}:

\begin{eqnarray}\label{fail}
	&& \hskip -1cm
	\Pr[\text{fail}]= \sum_{\tau+\beta \notin \Big(\frac{2^{t-2}}{4};\frac{(3\cdot 2^{t-2})}{4} \Big)} P_{\tau+\beta}\Big(t-\frac{2^{t-2}}{2}\Big) + \nonumber \\ &&
	\sum_{\tau+\beta \in \Big(\frac{2^{t-2}}{4};\frac{(3\cdot 2^{t-2})}{4} \Big)} P_{\tau+\beta}\Big(t\Big).
\end{eqnarray}

In lattice-based cryptography, the parameter $\sigma$ is often chosen significantly smaller than the modulus $\rho=2^{t-1}$. The ratio $\sigma \ll \frac{p}{8}$ is common to prevent wrap-around modulo $\rho$ and to maintain hardness assumptions \cite{MS_err_corr,Ring-LWE}. Therefore, below in Table \ref{tab:failure-all} we present the probability of failure results for selected values of the discrete Gaussian distribution defined in $\mathbf{Z}_\rho$ parameter $\sigma = \rho^{1/4}$ and $r$ equations. 

\begin{table}[h]
		\centering
		\begin{tabular}{|c|c|c|c|}
			\hline
			\rule{0pt}{2.5ex} 
			$r$ & $\rho=2^{10}$ & $\rho=2^{11}$ & $\rho=2^{12}$ \\
			\hline
			$16$ & 4.8e-28  & 3.78e-78 & 1.13e-220 \\
			\hline
			$20$ & 1.7e-22 & 1.81e-62 & 2.53e-176 \\
			\hline
			$25$ & 5.0e-18 & 6.44e-50 & 7.72e-141 \\
			\hline 
			$30$ & 4.7e-15 & 1.53e-41 & 3.54e-117 \\
			\hline
			$50$ & 4.4e-09 & 9.0e-25  & 7.88e-70 \\
			\hline
			$100$ & 1e-4    & 3.73e-12 & 2.83e-34 \\
			\hline
			$200$ & 0.03    & 7.85e-06 & 1.8e-16 \\
			\hline
		\end{tabular}
	\caption{Probability of failure for different values of $r$, $\rho = 2^{t-1}$, $\sigma = \rho^{1/4}$}
	\label{tab:failure-all}
\end{table}

We can see from \Cref{tab:failure-all} that the probability of error increases with the number of equations. However, we can see that the failure probability remains acceptably low given that proper parameter values are chosen. As $\rho$
increases, a sharp decrease in the error probability is observed, for fixed $r$. Based on the performed calculations, we can estimate the dependency of the failure probability on $\rho$ by an exponential decay in $\rho$:
\[
	\Pr[\text{fail}]= A(r) \cdot e^{-B(r) \cdot \rho}
\]

where $ A(r) $ and $ B(r) $ are parameters depending on $r$. Due to this exponential decay we claim that with an appropriately chosen value of \(\sigma\), the probability of incorrect decryption becomes negligibly small. This completes the proof.
\end{proof}

\section{Security analysis}

We now consider the security of the presented scheme. 

Inspired by LWE hardness assumption (see  Section 2.6 in \cite{ABB2010}) we define two oracles $\mathcal{O}_{enc}$ and $\mathcal{O}_{rand}$ whose behaviors are presented below:

\begin{itemize}
	\item $\mathcal{O}_{enc}$ outputs samples $(\vec{w}_i, \varepsilon_i \cdot \vec{w}_i^{\vec{x}})$, where the entries of vectors $\vec{w}_i$ are uniformly sampled from one of the cycles as presented in \Cref{m16_2c_PKE} with a fixed value of $n_c$, i.e. for any query index $i$ and for a fixed column index $j \leq n_c$ we have $w_{ij} \in \langle ba \rangle$, otherwise $w_{ij} \in \langle a \rangle$. As previously, the vector $\vec{x}$ is uniformly chosen from $\mathbf{Z}_{2^t}^m$. The error $\varepsilon_i$ is an element of $\langle ba \rangle$.
	\item $\mathcal{O}_{rand}$ generates random samples of the form $\vec{w}_i, v_i$, where $\vec{w}_i$ is generated as above and $v_i \in \mathbf{M}_{2^t}$.
\end{itemize}

Defining the advantage of $\mathcal{A}$ to decide the LWE problem $LWE1(\mathbf{M}_{2^t}, m)$ presented above as
\begin{equation}
	AdvLWE1(\mathbf{M}_{16}, m) = |\Pr[\mathcal{A}^{\mathcal{O}_{enc}} = 1] - \Pr[\mathcal{A}^{\mathcal{O}_{rand}} = 1]|
\end{equation}
we show that the following proposition holds:

\begin{proposition}\label{prop:main_sec}
	The \Cref{m16_2c_PKE} oracle $\mathcal{O}_{enc}$  produces samples statistically indistinguishable from the ones produced by $\mathcal{O}_{rand}$.
\end{proposition}

\begin{proof}
Due to the specific structure of matrix $W$ each product $\vec{w}_i^{\vec{x}}$ can be divided into two parts $\omega_1$ and $\omega_2$, such that  $\omega_1 \in \langle ba \rangle$ and $\omega_2 \in \langle a \rangle$ just as in the previous proof. Moreover, both $\omega_1$ and $\omega_2$ are uniformly distributed in the appropriate cycles, since entries of $W$ and secret key $x$ were sampled uniformly from the appropriate sets.  Note that the error $\varepsilon_i$ can be viewed as a part of $\omega_1$. We can now construct a multiplication table for the value of the product $\omega_1 \cdot \omega_2$ to show that each element of $\mathbf{M}_{2^t}$ is equally likely, i.e. for a pair of two random elements $\omega_1 \in \langle ba \rangle$ and $\omega_2 \in \langle a \rangle$ the probability $\Pr[\omega_1 \cdot \omega_2 = \omega_0] = 2^{-t}$ for any fixed value $\omega_0 \in \mathbf{M}_{2^t}$. This completes the proof, since the oracle $\mathcal{O}_{rand}$ produces uniformly distributed $v_i$'s as well.
\end{proof}

Note that due to Bob's errors the element $v^\prime$ is of the form $ba^k$ for some $k$. Therefore, this element has a special structure which is an important fact to keep in mind for the security analysis. In other words, we can define the following Attack Game between the adversary $\mathcal{A}$ and the challenger $\mathcal{C}$:

\begin{enumerate}
	\item $\mathcal{C}$ creates an LWE instance as presented in this paper and sends the pair $(W, \vec{v})$ to $\mathcal{A}$;
	\item $\mathcal{A}$ samples a random bit $\mu \in \{0, 1\}$, and sends it to $\mathcal{C}$;
	\item $\mathcal{C}$ samples a random bit $\beta \in \{0, 1\}$;
	\item If $\beta = 0$, then $\mathcal{C}$ encrypts the bit $\mu$ by executing Bob's steps. Hence $\mathcal{C}$ obtains a ciphertext $\vec{c}_0 = (\vec{w}, a^{\beta \cdot 2^{t-2}} v^\prime)$;
	\item If $\beta = 1$, then $\mathcal{C}$ calculates $\vec{w}$ as Bob does, samples a random $k$ uniformly from the set $\left[0, 2^{t-1}\right]$ and outputs $\vec{c}_1 = (\vec{w}, ba^k)$;
	\item $\mathcal{C}$ sends $\vec{c}_{\beta}$ to $\mathcal{A}$;
	\item $\mathcal{A}$ outputs a guess $\hat{\beta}$. He wins if $\hat{\beta} = \beta$.
\end{enumerate}

Defining the advantage of $\mathcal{A}$ win the Attack Game presented above as
\begin{equation}
	AdvLWEenc(\mathbf{M}_{2^t}, m) = |\Pr[\hat{\beta} = \beta] - \frac{1}{2}|
\end{equation}
we show that the following proposition holds:

\begin{proposition}\label{prop:sem_sec}
	$\mathcal{A}$'s advantage in winning the Attack Game presented above is negligible.
\end{proposition}

\begin{proof}
	Since the entries of $\vec{v}$ are distributed uniformly, Bob's errors only place these entries in an appropriate set as presented in \Cref{m16_2c_PKE}. Therefore, powers of the generator $a$ are uniformly distributed in the set $\left[0, 2^{t-1}\right]$. Furthermore, the set $\mathbf{S}$ is chosen at random as well. Since only the first entry of $v^\prime$ contains the generator $b$, it cannot be eliminated and no additional factors appear during encryption. Hence, we can consider the distribution of the power of $a$ in $v^\prime = \prod_{i = 1}^{r} v^\prime_{s(i)}$. However, since $\mathbf{Z}_{2^{t-1}}$ is an additive group, the uniform distribution is preserved for the sum of powers of $a$. On the other hand, the element $ba^k$ from $\vec{c}_1$ was generated using a uniformly distributed $k$. Therefore, the adversary cannot distinguish between $a^{\beta \cdot 2^{t-2}} v^\prime$ and $ba^k$. 
\end{proof}

As for the recovery of the secret key, we have seen that the non-commutativity of $\mathbf{M}_{2^t}$ cannot be ignored. However, due to definitions \eqref{m16mult} and \eqref{m16exp} of the basic operations, $LWE(\mathbf{Z}_2, m)$ is a subproblem of $LWE1(\mathbf{M}_{2^t}, m)$.  Therefore, we claim that
\[
	AdvLWE1(\mathbf{M}_{2^t}, m) \leq AdvLWE(\mathbf{Z}_{2}, m).
\]

Note, however, that solving $LWE(\mathbf{Z}_2, m)$ would reveal the parity of the first $n_c$ entries of the secret vector $\vec{x}$. As of now, we do not know how this fact affects the security of our proposal. We think, this issue can be dealt with by replacing $2$ with a prime $q$ in the definition of  modular-maximal cyclic groups, i.e. in the future we may also consider a family of groups $\mathbf{M}_{q^t}$.

On the other hand, the LWE problem in $\mathbf{M}_{2^t}$ is not equivalent to the one in $\mathbf{Z}_\rho$, due to the non-commutativity of $\mathbf{M}_{2^t}$. The difference between two problems comes mainly from definition \eqref{m16exp}. This is an extra factor for the adversary $\mathcal{A}$ to consider in his search for the secret key.

\section{Comparison to the original LWE}

Let us now compare the presented schemes to Regev's original idea.

The first major difference is that the \Cref{m16_2c_PKE} uses two cycles of $\mathbf{M}_{16}$ in such a way that error terms must exist for it to work properly. In other words, the \Cref{Regev_PKE} without errors is easily broken using Gaussian elimination technique whereas in the \Cref{m16_2c_PKE} equations without errors are worthless, since the scheme is not able to decrypt a bit correctly (at least, not with an overwhelming probability). In other words, at least Bernoulli error terms are necessary for the scheme to function correctly. Also, since the non-commutativity of $\mathbf{M}_{2^t}$ cannot be ignored, even without errors the complexity of solving a system of equations obtained from the \Cref{m16_2c_PKE} is not known thus far. We think that this fact beneficially distinguishes our idea from the original one.

Let us now consider the original idea of Regev. Assume that the public matrix $A$ has a left pseudo-inverse matrix $A^\prime$, i.e. $A^\prime A$ is an identity matrix. Then the LWE problem $Ax + e = b$ can be restated as follows by multiplying the original problem by $A^\prime$ on the left:
\[
	x + A^\prime e = A^\prime b,
\]
where now $x$ serves as an error vector and $e$ is the private key. This is called the duality of LWE problem. In other words, the secret key $x$ and the error term $e$ are interchangeable in the original idea.

\begin{figure}
	\centering
	\includegraphics[width=0.85\textwidth]{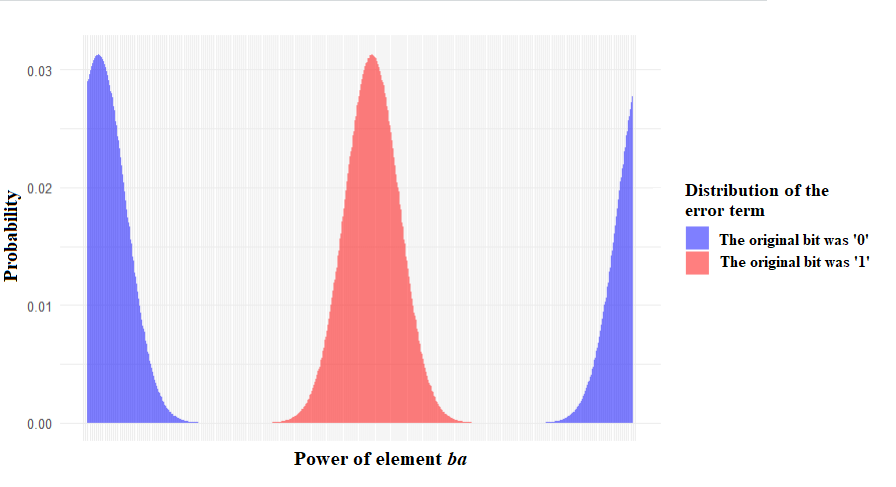}
	\caption{Shifted distributions of the error term $h$.}
	\label{fig:err}
\end{figure}

This property is lost in our scheme, since the discrete logarithm cannot be applied to the considered equations. Furthermore, the are two types of errors used in the scheme. As a consequence we obtain shifted distributions for the error term $h = \prod_{i=1}^r \delta_i \prod_{i=1}^r \varepsilon_i$. An example of these distributions for the parameters $\rho = 256$, $\sigma = 4$, $r = 10$ is presented in \Cref{fig:err}.

From the point of view of security, we think that both schemes presented here have at the very least a comparable level of security, considering that the size of the non-commuting group can be increased to any power of a prime $q$. This comes from the fact that all of the schemes rely on similar systems of equations. Regev's scheme relies on a classic system of linear equations with errors. The direct multiplicative approach is essentially the same after using a discrete logarithm mapping. However, as opposed to the direct approach the discrete logarithm mapping cannot be applied to LWE problem defined in non-commuting group $\mathbf{M}_{2^t}$, since this group itself brings an additional randomness factor into LWE problem.  

\section{Conclusions}

In this paper, we have presented a way for adapting the LWE problem to non-commuting multiplicative groups. In the presented scheme we consider a group with two non-commuting generators which cannot be considered separately, since in definitions \eqref{m16mult}, \eqref{m16exp} and \eqref{m16inv} the exponent of generator $b$ also affects the exponent of generator $a$ in the output value. We think that this fact is beneficial for our scheme, since the additive group $\mathbf{Z}_q$ used in the original LWE is cyclic generated by $1$.

Notably, the non-commuting group $\mathbf{M}_{2^t}$ contains two cycles of size $2^{t-1}$. We used this fact to show that it can be used to construct an accurate criterion for the correct decryption of a bit with overwhelming probability. We have proven that the ciphertext produced by the general concept scheme oracle $\mathcal{O}_{enc}$ is indistinguishable from a truly random output of the oracle $\mathcal{O}_{rand}$. 

The security of our scheme is at the very least comparable to the security of the original scheme by Regev. For now it is an open question how the non-commutativity factor contributes to the overall security of the scheme. However, we think that the original security can be easily achieved simply by switching the group $\mathbf{M}_{2^t}$ for a larger one, namely $\mathbf{M}_{p^t}$. In this paper, we focused more on establishing a working scheme leaving larger groups for future exploration.

\begin{credits}
\subsubsection{\ackname} This research received no funding. 
\subsubsection{\discintname}

The authors have no competing interests.

\end{credits}
%
%
%
%

\end{document}